\documentclass[aps,prl,twocolumn,superscriptaddress,showpacs]{revtex4-1}
\usepackage{graphicx}
\usepackage{bm}
\usepackage{color}
\usepackage{hyperref}

\begin{document}
\title{Eigenvalue statistics of reduced density matrix during driving and relaxation}
\author {M. Mierzejewski}
\affiliation{Institute of Physics, University of Silesia, 40-007 Katowice, Poland}
\author {T. Prosen}
\affiliation{Faculty of Mathematics and Physics, University of Ljubljana, SI-1000 Ljubljana, Slovenia }
\author {D. Crivelli}
\affiliation{Institute of Physics, University of Silesia, 40-007 Katowice, Poland}
\author{P. Prelov\v{s}ek}
\affiliation{Faculty of Mathematics and Physics, University of Ljubljana, SI-1000 Ljubljana, Slovenia }
\affiliation{J. Stefan Institute, SI-1000 Ljubljana, Slovenia }

\begin{abstract}
We study a subsystem of an isolated one-dimensional correlated metal when 
it is driven by a steady electric field  
or when it relaxes after driving. We obtain numerically exact reduced density matrix $\rho$ for subsystems which are
sufficiently large to give significant eigenvalue statistics and spectra of $\log(\rho)$. We show that both for generic 
as well as  for the integrable model the statistics follows  the universality  of Gaussian unitary and 
orthogonal ensembles for driven and equilibrium systems, respectively.  Moreover, the spectra of modestly driven subsystems are well described by the 
Gibbs thermal distribution with the entropy determined
by the time-dependent energy only.
%and the orthogonal quenched system  
%Quite unexpectedly this result is very robust and holds also for integrable systems as well as for states which are far from equilibrium.  

\end{abstract}
\pacs{71.27.+a,72.10.Bg,05.70.Ln}
%71.27.+a Strongly correlated electron systems; heavy fermions
%72.10.Bg General formulation of transport theory
%05.70.Ln 	Nonequilibrium and irreversible thermodynamics 
\maketitle

\textit{Introduction.}--- 
Spectral universality is one of the key features of highly excited complex systems. It has been 
demonstrated and observed in a diverse range of phenomenologies, ranging from acoustics \cite{ellegaard}, 
microwave resonators \cite{stoeckmann,*stoeckmann1}, quantum dots \cite{marcus}, to many-particle systems, 
such as complex nuclei \cite{bohigas} and strongly correlated models of condensed matter \cite{montambaux}. 
Universality is quantitatively characterized by the applicability of a parameter free random matrix theory (RMT) 
\cite{mehta}, where the Hermitian operator in question, usually the Hamiltonian, is described by an ensemble 
of Gaussian random matrices, where the only constraint,
 whether matrices are real symmetric, or complex hermitian, 
is imposed by the existence, or non-existence of a (generalized)
 time-reversal symmetry. Random matrix distribution of energy levels 
is also widely used as a clean indicator of complexity, or non-integrability,
 of a physical model, and is the most abstract definition of a quantum chaotic
 behavior \cite{hakke}.

In this Letter we propose RMT analysis of a completely different concept in quantum statistical physics, 
namely of spectra of reduced density matrix (RDM) $\rho$ of equilibrium and non-equilibrium states.
We consider RDM of strongly correlated quantum systems. In particular we study the 
one-dimensional (1D) model of interacting spinless fermions (equivalent to a Heisenberg--type spin chain)
for a variety of simple pure states of the entire system: the so-called microcanonical (MC) states 
(approximate eigenstates), the
time-evolving states after a quench of magnetic flux, or during inductive driving with a linearly increasing 
magnetic flux. We show that, quite remarkably, the statistics of eigenvalues of RDM of 
large subsystems is typically described by RMT. For equilibrium thermal states, we find agreement 
with Gaussian orthogonal ensemble (GOE), whereas for non-equilibrium, driven states, with currents, 
we find agreement with Gaussian unitary ensemble (GUE). We note in particular, that spectra of
RDM of large subsystems typically follow RMT even if the entire system is completely integrable.
 
Furthermore, the RDM can serve as a stringent test of thermal properties of 
nonequilibrium states and their thermalization. Our results show that for modestly driven systems 
the entropy density $s$  of the subsystem develops in time according to the quasi-equilibrium scenario,
i.e.  $s$ depends only on the instantaneous energy density $\varepsilon$. 
In agreement with the latter
is the observation for {\em driven integrable} and {\em non--integrable} chains 
that the eigenvalue spectra are consistent with the canonical Gibbsian form 
$\rho \propto \exp(-H_{\rm eff}/T)$ with well defined effective temperature $T$ and
 $H_{\rm eff}$ being a $T-$independent effective Hamiltonian of the subsystem.

\textit{Model and method.}--- 
We study the 1D model of interacting spinless fermions on a chain of even number of sites $L$
with periodic boundary conditions. We investigate how the system responds to an external electric
field as introduced in the time-dependent model by the varying magnetic flux  $\phi(t)$, 
\begin{eqnarray}
H(t)&=&-t_0 \sum_j \left\{ {\mathrm e}^{i \phi(t)/L}\; c^{\dagger}_{j+1}c_j +{\mathrm h.c.} \right\} \nonumber  \\
& &+ V \sum_j \hat{n}_j \hat{n}_{j+1} +W \sum_j \hat{n}_j \hat{n}_{j+2},
\label{ham}
\end{eqnarray}
where $\hat{n}_j= c^{\dagger}_{j}c_j$, $t_0$ is the hopping integral, 
$V$ and $W$ are the repulsive potentials between fermions on the nearest-neighbor and the 
next-nearest-neighbor  sites. Furtheron we use units in which $\hbar=k_B=t_0=1$.
The main idea behind introducing $W$ is to break integrability of the pure $t_0-V$ model.
One expects generic properties for the non--integrable case with $W \ne 0$, whereas the integrable system
($W=0$) shows anomalous relaxation \cite{gge,Eckstein2012,Cassidy2011,my3} and transport characteristics \cite{u2,my1,my2,Tomaz2011,Marko2011,Sirker2009,Robin2011}.   
If not stated otherwise,  the numerical results for integrable and non--integrable cases will 
refer to $V=1$, $W=0$ and $V=1.4$, $W=1$ systems, respectively, 
at half-filling with $M=L/2$ fermions on $L=26$ sites. These parameters correspond to the
metallic regime.  

In the numerical procedure using the microcanonical (MC) Lanczos method \cite{mclm}
we generate initial states $|\Psi(0) \rangle$  for the target energy $E_0=\langle \Psi(0)| H(0) |\Psi(0) \rangle$ and 
the energy uncertainty $\delta^2 E_0=\langle \Psi(0)| [H(0)-E_0]^2|\Psi(0) \rangle $. 
Typically, we consider large $E_0$ (corresponding to high $T$, i.e. $\beta\equiv 1/T <0.5$) with $L=26$ and 
$\delta \simeq 0.01$. To simulate the MC ensemble, the energy window 
is small on a macroscopic scale ($\delta E_0/E_0 \ll 1$) but still contains a large number of levels \cite{Goldstein2006}. 
The time evolution of $ |\Psi (0) \rangle \rightarrow |\Psi (t) \rangle$ 
is calculated by  Lanczos propagation method \cite{lantime} 
applied to small time intervals $(t,t+\delta t)$. 

Since the entire chain is isolated from the surroundings it remains in a pure state 
$|\Psi (t) \rangle \langle \Psi (t)|$. In this Letter we focus on the reduced dynamics 
of its subsystem containing $N$ subsequent lattice sites. The RDM of the subsystem is
then $\rho=\mathrm{Tr}_{L-N} |\Psi (t) \rangle \langle \Psi (t)|$ where the trace is taken over the remaining
$L-N$ sites.   RDM is block--diagonal  with respect to number of particles in the subsystem  
$n =\sum_{i=1}^{N} n_i = {0,...,N}$. As the approach is numerically accurate the reduced dynamics 
is exact as well.   At the same time,  the approach  allows for subsystems which are sufficiently 
large to give meaningful level statistics as well as spectral and other properties of $\rho$. 
With respect to time-dependent response we study two kinds of systems: 
a) driven by a steady electric field $F=-\partial_t \phi(t)/L=\mathrm{const}$, and b) 
relaxing but not necessarily thermalizing after  a sudden quench of the flux (field pulse): 
$\phi(t)=\theta(t)\delta \phi$. 

\textit{Eigenvalues statistics of RDM}--- 
We start with the presentation of results on the eigenvalue statistics of $\log(\rho)$.
%Later on, we substantiate our claims concerning the regimes 
%when  the subsystem approaches the thermal,  steady non--thermal or nonequilibrium states. 
For the quasi--thermal states
$\rho \propto \exp(-\beta H_{\rm eff})$
the statistics of eigenvalues of $\log(\rho)$  should be the same as the level statistics of the effective Hamiltonian.
Since we can reach using MC Lanczos method (at half-filling) systems with $L=26$ 
the  eigenvalue statistics is
determined for the largest accessible subsystems of $N=12$ sites  and 
$n=6$ fermions (containing  924 levels).
 We note that even though the number $n$ of fermions within the subsystem 
is not conserved, RDM is block-diagonal with respect to states with fixed $n$, 
hence $[H_{\rm eff},n]=0$.
The spectrum $\left\{ \lambda_i \right\}$  of $\log(\rho)$ is unfolded by a linear interpolation of the 
integrated density ${\cal N}(\omega)=\sum_i \theta(\omega-\lambda_i) $ in intervals containing 
$40$ subsequent eigenvalues. This procedure leads to a smoothened integrated density 
${\cal N}_{\mathrm sm} (\omega)$.   The unfolded spectrum consists of
$\bar{\lambda}_i={\cal N}_{\mathrm sm} (\lambda_i)$ where we analyze only $~2/3$ of
eigenvalues from middle of this spectrum (see \cite{Bruus1997} for more details on unfolding). 
We study in the following two standard quantities characterizing the level statistics:
a) the nearest-level-spacing distribution $p(x)$ and the eigenvalue 
number variance $\Sigma^2(\delta)$ defined as the variance of the number of unfolded 
eigenvalues in the interval of length $\delta$. 
%$p(x)$ is the probability density for two neighboring eigenvalues having the  spacing $(\bar{\lambda}_{i+1}-\bar{\lambda}_i)$ equal $x$.$\Sigma^2(\delta)$  is 
$p(x)$ and  $\Sigma^2(\delta)$ characterize local correlation properties of the spectrum and 
long-range level correlations, respectively. 
The numerical results for $\log(\rho)$ can be compared with the results  of the RMT for the GOE or GUE
ensembles \cite{Brody1981,Guhr1998},
\begin{eqnarray}
p_{\mathrm{GOE}}(x)&\simeq & \frac{\pi x}{2} \exp(-\pi x^2/4), \label{goep}  \\
\Sigma^2_{\mathrm{GOE}}(\delta) &\simeq &  \frac{2}{\pi^2} \left( 
\log(2\pi \delta)+\gamma+1-\frac{\pi^2}{8} \right) , \label{goes}  \\
p_{\mathrm{GUE}}(x)&\simeq & \frac{32 x^2}{\pi^2} \exp(-4 x^2/ \pi), \label{guep} \\
\Sigma^2_{\mathrm{GUE}}(\delta) &\simeq &  \frac{1}{\pi^2} \left( 
\log(2\pi \delta)+\gamma+1\right),  \label{gues}
\end{eqnarray}
where $\gamma \simeq 0.577 $ is the Euler constant. 
Note that Hamiltonians of many-body integrable systems have the 
Poisson distribution with $p_{\mathrm{P}}(x)=\exp(-x)$ and 
$\Sigma^2_{\mathrm{P}}(\delta)=\delta$, while generic non-integrable systems with the time--reversal 
symmetry are expected to follow the GOE statistics. Only cases breaking the time-reversal symmetry 
should result in the GUE statistics.

\begin{figure}
\includegraphics[width=0.4\textwidth]{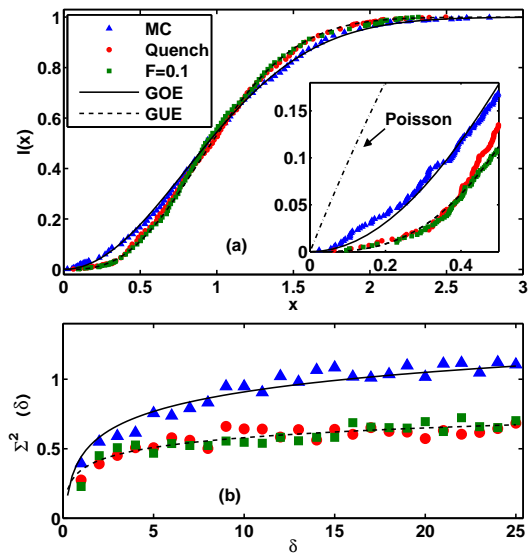}
\caption{(Color online) Eigenvalue statistics of  $\log(\rho)$ of the integrable system: in the equilibrium MC state, 
in the quasi-equilibrium evolution during driving ($F=0.1$), and in the non--thermal steady state after 
quenching the flux (Quench).  Panel (a) shows the integrated spacing distribution $I(x)$  with the inset 
zoom in  the low--gap regime, and b) the level number variance $\Sigma_2(\delta)$. 
Lines show results from  the RMT.}
\label{fig1}
\end{figure}

%We have initially studied the most obvious cases of generic non--integrable systems in the quasi--thermal state \cite{Goldstein2006,Riera2012}
%when  $\rho \propto \exp[-\beta H_{eff}(\phi)]$. 
 We first analyze the spectrum of $\log(\rho)$ for the initial MC state in a generic non--integrable system
and find that both $p(x)$ 
and $\Sigma^2(\delta)$  accurately reproduce the results for GOE (not shown).  It is  
expected since in this case without a flux the time-reversal symmetry  is preserved and 
$\rho$  can be chosen as a real symmetric matrix. 
In Fig.~\ref{fig1} we present numerical data for the integrable system together with the prediction of the RMT. 
The upper panel shows integrated spacing distribution $I(x)=\int^x_0 \mathrm{d}y p(y)$, whereas  
$\Sigma^2(\delta)$ is shown in the lower panel. Surprisingly, the eigenvalue statistics of the RDM 
of a subsystem turns out to be independent of the integrability of the total system.  

On the other hand, under a constant (but modest) field $F > 0 $ \cite{my1,my2}  
or after a sudden flux quench \cite{Rigol2012} we find that the statistics turn into GUE. This is the case for the
non--integrable systems as well as for the integrable one, as clearly confirmed 
in Fig.~1.   The GUE statistics at $F>0$ is consistent with the time--symmetry breaking 
by a finite current within the subsystem. In the case of quenching the decay of the current is  not complete, 
at least not within an integrable system where the absence of the current relaxation is a hallmark of 
a finite charge stiffness \cite{u2}.  
%It is in order to recall that 
%the integrable system relaxes after quenching to non--thermal steady state, when $\rho$ may be very different from 
%$\exp[-\beta H_{eff}(\phi)]$.  Still the eigenvalue statistic stays within the GUE as shown in Fig.~\ref{fig1} 
%(see points labelled as ``Quench'').
We have determined the eigenvalue statistics also for the far-from-equilibrium driven states, shortly after 
the electric field has been switched on (not shown).  The eigenvalues of $\log(\rho)$
are very similar to those presented in Fig.~\ref{fig1} for the GUE level case,
indicating that  the RMT statistics of  $\log(\rho)$ is very robust. 
% and holds also far beyond the
%quasi-thermal regimes, where it might be expected.}

\begin{figure}
\includegraphics[width=0.48\textwidth]{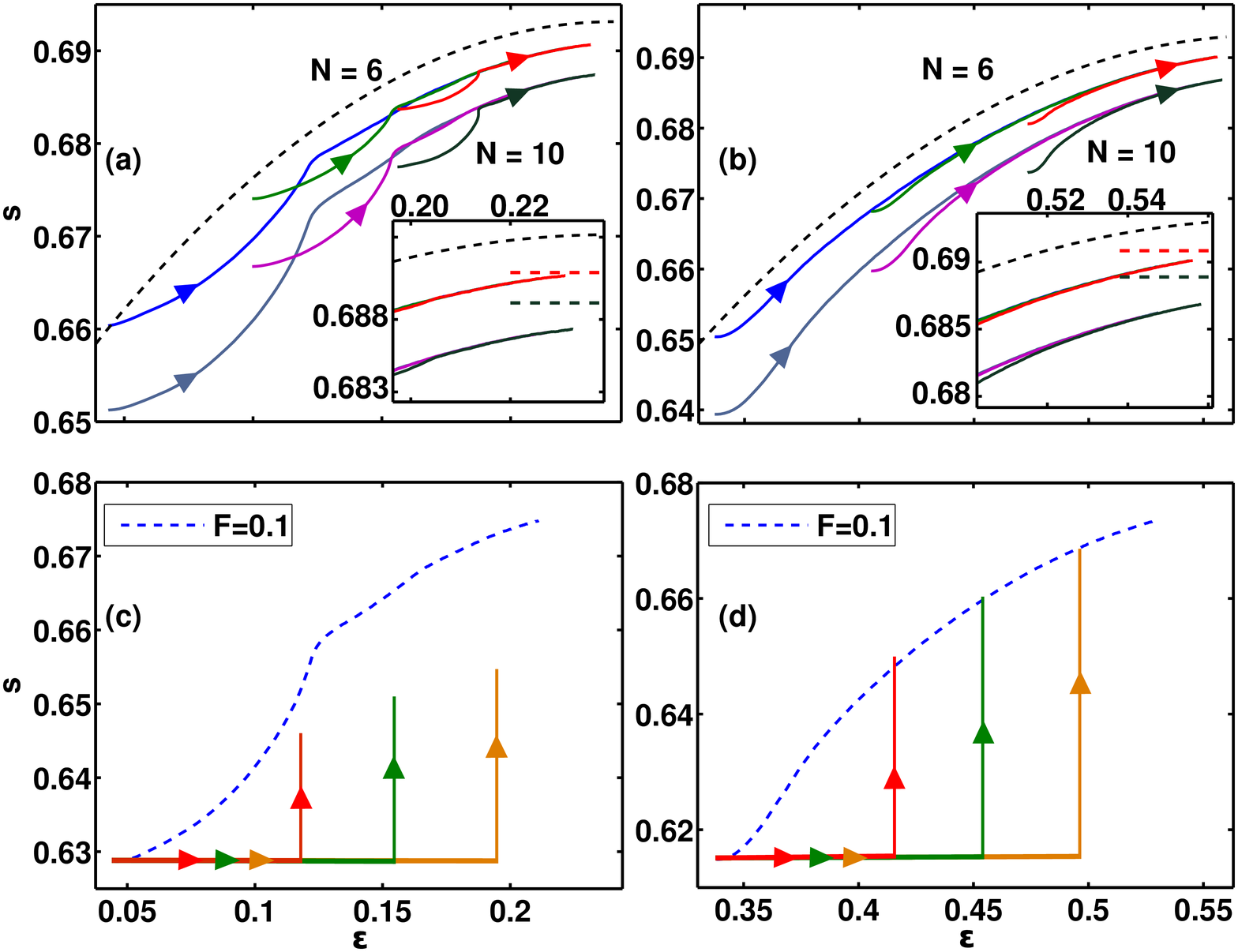}
\caption{(Color online) 
Instantaneous entropy density $s(t)$ vs. energy density $\varepsilon(t)$ 
for integrable (a,c) and non--integrable (b,d) metals.
Panels (a,b) show results for systems driven by field $F=0.1$ and
$N=6,10$ subsystems. Dashed curves represent high--$T$ analytic results, Eq.~(\ref{htes}) for $L \rightarrow \infty$. 
In the insets the  high-$\varepsilon$ regime is magnified and $1/L$ corrections
to $s_{\infty}$ are included  and marked as horizontal lines.
Panels (c,d) show relaxation after flux quenching for $N=12$ subsystem. 
Here, dashed curves show quasi-equilibrium results for the same systems driven with  $F=0.1$.
Arrows mark the direction of the processes.  }
\label{fig2}
\end{figure}

\textit{Subsystem entropy density}- While the von Neumann entropy of the total 
system is at all times zero for the pure state  $|\Psi (t) \rangle \langle \Psi (t)|$, the (entanglement)  
entropy of the subsystem  $S=-\mathrm{Tr}_N(\rho \log \rho)$ is clearly not.  
As a strong  indication of the  thermal (or quasi-equilibrium) states we can use
 the relation of the subsystem entropy density  $s=S/N$ and the energy density 
 $\varepsilon=\langle H(t) \rangle/L$ of the total system (being the same for the subsystem).
Hence, we present in Fig.~\ref{fig2}  the time-evolution of $s(t)$ plotted versus 
$\varepsilon(t)$ for two systems: driven by a constant 
electric field $F=0.1$ (panels a,b) and after a sudden flux quench (panels c,d). 
Results in the right and left panels are
for the non--integrable and integrable  cases, respectively.  
Note first that $s(t)$ only weakly depends on $N$ confirming its macroscopic relevance \cite{rigolast}. 
This is in  contrast with a specific case of the
ground state where we have found $s(0) \propto N^{-1}$ in agreement with 
the so called area laws for the entanglement entropy \cite{Eisert2010}.

Since we are studying the high-$T$ regime it is instructive to recall the equilibrium 
result following from straightforward high temperature expansion (HTE),  
\begin{eqnarray}
\varepsilon&=& \varepsilon_{\infty} -\beta \sigma^2_{\infty},   \label {hte} \\
\varepsilon_{\infty} & \simeq & \frac{V+W}{4}(1-\frac{1}{L}),  \label {hte1} \\
\sigma^2_{\infty}  & \simeq & \frac{t^2}{2}+\frac{V^2+W^2}{16}+\frac{t^2}{2L}-\frac{VW}{4L} \label {hte2} 
\end{eqnarray}
where $\varepsilon_{\infty}, \sigma_{\infty}$ 
refer to $T=\infty$ and $1/L$ corrections emerge due to a
restriction of strictly $M=L/2$ fermions in the whole system. Integrating the  equilibrium  relation
 $\beta=\partial s/\partial \varepsilon$ one gets for the equilibrium entropy density
\begin{eqnarray}
s(\varepsilon)&=&s_{\infty}-\frac{( \varepsilon_{\infty}-\varepsilon)^2}{2 \sigma^2_{\infty}}
\label{htes} \\
s_{\infty} &\simeq& \log(2)-\frac{N-1}{4L^2}-\frac{N^2-1}{6L^3}, \label{sinf}
\end{eqnarray}
where again leading $1/L$ corrections in $s_\infty < \log(2)$ arise due to fixing
$M=L/2$. Insets in Figs.~\ref{fig2}a,b  show that our numerical results 
are very close to this simple estimate for $s_{\infty}$.

The relation (\ref{htes}) allows specifying regimes which are clearly nonequilibrium or steady but non--thermal. 
The former case occurs, e.g., just after turning on the electric field when  
$s=s(\varepsilon)$ is convex (see Figs.~\ref{fig2}a and \ref{fig2}b) contrary to concave dependence, 
which according to Eq.~(\ref{htes}) should characterize the quasi--equilibrium evolution. 
More interesting is the observation in Fig.~\ref{fig2}c that the stationary non--thermal state emerges 
when integrable system relaxes after a sudden quench \cite{gge,Eckstein2012} but $s(t)$ 
remains evidently smaller than expected for a thermal relation, Eq.~(\ref{htes}). 
%\marcin{This result 
%nicely complies with the  hypothesis of the generalized Gibbs ensemble (GGE) since 
%$\rho_{GGE}$ maximizes the 
%entropy but only subject to constraints imposed by the integrals of motion \cite{gge,Cassidy2011}. }

Results shown in Fig.~\ref{fig2} may also suggest regimes, when the system even out of equilibrium 
reaches a quasi--thermal state. In this case $s(\varepsilon)$ should become independent of the 
initial state and close to the prediction of HTE. This indeed happens for integrable or non--integrable systems 
driven long enough by a moderate  steady $F$ (note that $F$ breaks integrability of an integrable system) 
or when non--integrable system relaxes after the flux  quench (see Fig.~\ref{fig2}d).

%Before investigating this quasi--thermal behavior in more details we briefly 
%explain why $s_{\infty} < \log(2)$. This finite--size effect has the same origin as  $1/L$ corrections in
%Eqs. (\ref{hte1} - \ref{hte2}) and  is due to the fixed  number 
%of particles in the total system.  
%In particular, for a subsystem of $N=2$ sites  at $\beta=0$ 
%one finds that  the states with $n=1$ occur with probabilities $p_{01}=p_{10}=\frac{1}{4}(1+\frac{1}{L})>\frac{1}{4}$, while probabilities for $n=0$ or $n=2$ are $p_{11}=p_{00}=\frac{1}{4}(1-\frac{1}{L}) < \frac{1}{4}$.
%Hence, $s_{\infty}= -\frac{1}{N}\sum_i p_i \log(p_i)<\log(2)$.  
%Calculations similar to HTE give the leading terms in $1/L$ expansion:

\begin{figure}
\includegraphics[width=0.48\textwidth]{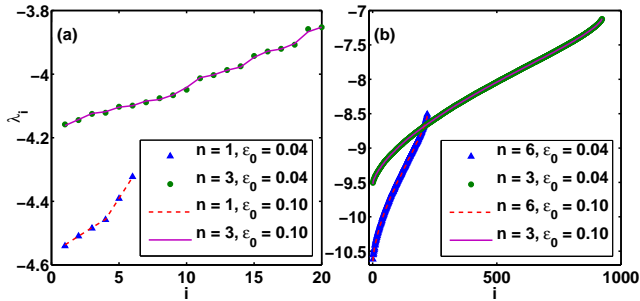}
\caption{(Color online) 
Spectra of $\log[\rho(t)]$ for the integrable system driven by field $F=0.1$ determined
at time $t$ such that $\varepsilon(t)=0.21$ but different initial energy densities
$\varepsilon_0=0.04, 0.1$ and different particle subsectors $n$. Panels (a) and (b) shows 
results for subsystems with $N=6$ and $N=12$ sites, respectively. }
\label{fig3}
\end{figure}

\textit{Thermal states}-
More stringent test for a thermal state is the requirement that the RDM obeys the 
canonical distribution  $\rho \propto \exp[-\beta H_{\rm eff}(\phi)]$ whereby $H_{\rm eff}$ plays the role
of an effective subsystem Hamiltonian. The effective inverse temperature $\beta$ can 
be also simply obtained from Eq.~(\ref{hte}).
The main open problem  concerns  the meaning of  $H_{\rm eff}(\phi)$ when
the subsystem is strongly coupled to its surroundings or
is subject to external driving.
However, we avoid this problem by testing the thermalization hypothesis without specifying 
the explicit form of $H_{\rm eff}(\phi)$.

In order to verify that the quasi--thermal state is determined solely by the 
 energy density,
we have determined the reduced dynamics of two identical subsystems (labelled by subscripts 1 and 2) 
driven by the same field $F=0.1$ but starting from different initial energies. 
We compare $\rho_1(t_1)$ and $\rho_2(t_2)$ for such times $t_1$ and $t_2$ that both systems
have the same instantaneous energies (temperatures) $E(t_1)=E(t_2)$.  
Then, one expects 
$\log[\rho_1(t_1)]=-\beta H_{\rm eff}[\phi(t_1)] +\mathrm{const}$ and 
$\log[\rho_2(t_2)]=-\beta H_{\rm eff}[\phi(t_2)] +\mathrm{const}$. 
In other words,  for the quasi thermal state operators $\log[\rho_1(t_1)]$ and $\log[\rho_2(t_2)]$ should give 
 Hamiltonians of the same system but at least subject to different fluxes. 
Such Hamiltonians  may have different eigenfunctions but the energy spectrum should be the same 
(up to eigenvalue fluctuations discussed above). 
%Although the operators  $\log[\rho_1(t_1)]$ and $\log[\rho_2(t_2)]$ may be different, their spectra should 
%be identical. 
Fig.~\ref{fig3} shows results testing the above hypothesis for two different subsystems $N=6,12$ and two sectors $n$,
respectively. It is quite evident that the spectra are independent of the initial MC energy density $\epsilon_0$, 
at least for the states for which the results on entropy 
already suggested possible thermalization.

\begin{figure}
\includegraphics[width=0.48\textwidth]{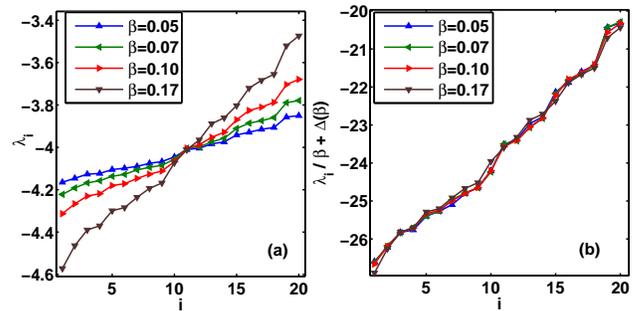}
\caption{(Color online) 
Spectra of $\log[\rho]$ for the integrable system ($N=6,n=3$) driven by $F=0.1$ determined for $t$
corresponding to different $\beta(t)$ obtained from HTE. Panels (a) and (b)
show eigenvalues of  $\log[\rho]$  and  normalized (and shifted) $\log[\rho]/\beta$, respectively. 
} 
\label{fig4}
\end{figure}

%The above test consists in comparing systems with the same energy but evolving 
%from different initial conditions. 
%One can also compare systems with different energies and 
%check the exponential dependence between $\rho$ and $\beta$. 
Fig.~\ref{fig4}a finally shows spectra of $\log [\rho(t)]$ for driven integrable system 
(only the largest sector with $n=3$ is presented).  
Various curves are obtained for various times of driving, 
when the system has different instantaneous energies $\varepsilon(t)$. 
Together with Eq.~(\ref{hte}) $\varepsilon(t)$ is used to determine the 
effective $\beta(t)$.  For a quasi--thermal evolution the spectra of 
$\log[\rho(t)]/\beta(t)=-H_{\rm eff}[\phi(t)]+\mathrm{const}$ should be the same up to a constant 
value.  As shown in Fig.~\ref{fig4}b, even the driven integrable system perfectly fulfills this 
requirement. The thermalization of non--integrable systems is commonly expected 
(apart from a few more specific cases \cite{Manmana2007,Gogolin2011}), 
and our results (not shown) confirm that in general.
%we have 
%focused on driven integrable system, when quasi--thermal evolution 
%is by far not obvious. Results for non--integrable cases are qualitatively similar.}    

\textit{Discussion.}--- 
 Our study shows that the RDM $\rho$ of a subsystem within the isolated system of interacting fermions
can be a useful tool to investigate properties of a nonequilibrium system.  Our results
on the RDM eigenvalue statistics within a 1D model of interacting spinless fermions 
reveal a universal conclusion, that subsystem of an equilibrium
MC state obeys the GOE eigenvalue statistics, independent of integrability or  non--integrability
of the whole  system (note that an integrable system as a whole obeys the
 Poisson statistics for the total energy eigenvalues). 
Moreover,  subsystems of the driven system and systems quenched with a field pulse 
follow the GUE universality, although the model by itself does not break the time--reversal symmetry.

Further, the spectrum of $\log(\rho)$  contains information useful for identifying the quasi--thermal,
steady non--thermal and the non--equilibrium regimes. 
For the case of quasi--thermal states, which
are realized also for finite but  modest driving, 
we have  demonstrated that the effective inverse temperature $\beta$ as the only relevant 
parameter determines the spectra of eigenvalues of $\log(\rho)$.  
On one hand, this result sets straightforward limits on the  relaxation 
of integrable systems.
But possibly more importantly,  
it introduces a nontrivial concept of subsystem's effective Hamiltonian $H_{\rm eff}$.
The physical content and the usefulness of the latter, also  its relation to the original full $H$ 
has still to be explored.  In any case it gives a novel approach to investigations of 
non--equilibrium properties of isolated interacting systems,
in particular in relation to their thermal and transport response.

%In the former case we were unable to check directly  whether the whole density matrix 
%is determined by a single parameter $\beta$, 

\acknowledgements

M.M. acknowledges support from the N202052940 project of NCN and the
ESF activity 'Exploring the Physics of Small Devices (EPSD)'. 
D.C. acknowledges a scholarship from the TWING project, co-funded by the European Social Fund.
T.P. and P.P. acknowledge the support by the Program P1-0044 and project J1-4244 of the Slovenian 
Research  Agency.
 
\bibliography{bibliography}

\begin{thebibliography}{30}%
\makeatletter
\providecommand \@ifxundefined [1]{%
 \@ifx{#1\undefined}
}%
\providecommand \@ifnum [1]{%
 \ifnum #1\expandafter \@firstoftwo
 \else \expandafter \@secondoftwo
 \fi
}%
\providecommand \@ifx [1]{%
 \ifx #1\expandafter \@firstoftwo
 \else \expandafter \@secondoftwo
 \fi
}%
\providecommand \natexlab [1]{#1}%
\providecommand \enquote  [1]{``#1''}%
\providecommand \bibnamefont  [1]{#1}%
\providecommand \bibfnamefont [1]{#1}%
\providecommand \citenamefont [1]{#1}%
\providecommand \href@noop [0]{\@secondoftwo}%
\providecommand \href [0]{\begingroup \@sanitize@url \@href}%
\providecommand \@href[1]{\@@startlink{#1}\@@href}%
\providecommand \@@href[1]{\endgroup#1\@@endlink}%
\providecommand \@sanitize@url [0]{\catcode `\\12\catcode `\$12\catcode
  `\&12\catcode `\#12\catcode `\^12\catcode `\_12\catcode `\%12\relax}%
\providecommand \@@startlink[1]{}%
\providecommand \@@endlink[0]{}%
\providecommand \url  [0]{\begingroup\@sanitize@url \@url }%
\providecommand \@url [1]{\endgroup\@href {#1}{\urlprefix }}%
\providecommand \urlprefix  [0]{URL }%
\providecommand \Eprint [0]{\href }%
\providecommand \doibase [0]{http://dx.doi.org/}%
\providecommand \selectlanguage [0]{\@gobble}%
\providecommand \bibinfo  [0]{\@secondoftwo}%
\providecommand \bibfield  [0]{\@secondoftwo}%
\providecommand \translation [1]{[#1]}%
\providecommand \BibitemOpen [0]{}%
\providecommand \bibitemStop [0]{}%
\providecommand \bibitemNoStop [0]{.\EOS\space}%
\providecommand \EOS [0]{\spacefactor3000\relax}%
\providecommand \BibitemShut  [1]{\csname bibitem#1\endcsname}%
\let\auto@bib@innerbib\@empty
%</preamble>
\bibitem [{\citenamefont {Ellegaard}\ \emph {et~al.}(1995)\citenamefont
  {Ellegaard}, \citenamefont {Guhr}, \citenamefont {Lindemann}, \citenamefont
  {Lorensen}, \citenamefont {Nyg\aa{}rd},\ and\ \citenamefont
  {Oxborrow}}]{ellegaard}%
  \BibitemOpen
  \bibfield  {author} {\bibinfo {author} {\bibfnamefont {C.}~\bibnamefont
  {Ellegaard}}, \bibinfo {author} {\bibfnamefont {T.}~\bibnamefont {Guhr}},
  \bibinfo {author} {\bibfnamefont {K.}~\bibnamefont {Lindemann}}, \bibinfo
  {author} {\bibfnamefont {H.~Q.}\ \bibnamefont {Lorensen}}, \bibinfo {author}
  {\bibfnamefont {J.}~\bibnamefont {Nyg\aa{}rd}}, \ and\ \bibinfo {author}
  {\bibfnamefont {M.}~\bibnamefont {Oxborrow}},\ }\href {\doibase
  10.1103/PhysRevLett.75.1546} {\bibfield  {journal} {\bibinfo  {journal}
  {Phys. Rev. Lett.}\ }\textbf {\bibinfo {volume} {75}},\ \bibinfo {pages}
  {1546} (\bibinfo {year} {1995})}\BibitemShut {NoStop}%
\bibitem [{\citenamefont {St\"ockmann}\ and\ \citenamefont
  {Stein}(1990)}]{stoeckmann}%
  \BibitemOpen
  \bibfield  {author} {\bibinfo {author} {\bibfnamefont {H.-J.}\ \bibnamefont
  {St\"ockmann}}\ and\ \bibinfo {author} {\bibfnamefont {J.}~\bibnamefont
  {Stein}},\ }\href {\doibase 10.1103/PhysRevLett.64.2215} {\bibfield
  {journal} {\bibinfo  {journal} {Phys. Rev. Lett.}\ }\textbf {\bibinfo
  {volume} {64}},\ \bibinfo {pages} {2215} (\bibinfo {year}
  {1990})}\BibitemShut {NoStop}%
\bibitem [{\citenamefont {Sridhar}(1991)}]{stoeckmann1}%
  \BibitemOpen
  \bibfield  {author} {\bibinfo {author} {\bibfnamefont {S.}~\bibnamefont
  {Sridhar}},\ }\href {\doibase 10.1103/PhysRevLett.67.785} {\bibfield
  {journal} {\bibinfo  {journal} {Phys. Rev. Lett.}\ }\textbf {\bibinfo
  {volume} {67}},\ \bibinfo {pages} {785} (\bibinfo {year} {1991})}\BibitemShut
  {NoStop}%
\bibitem [{\citenamefont {Marcus}\ \emph {et~al.}(1992)\citenamefont {Marcus},
  \citenamefont {Rimberg}, \citenamefont {Westervelt}, \citenamefont
  {Hopkins},\ and\ \citenamefont {Gossard}}]{marcus}%
  \BibitemOpen
  \bibfield  {author} {\bibinfo {author} {\bibfnamefont {C.~M.}\ \bibnamefont
  {Marcus}}, \bibinfo {author} {\bibfnamefont {A.~J.}\ \bibnamefont {Rimberg}},
  \bibinfo {author} {\bibfnamefont {R.~M.}\ \bibnamefont {Westervelt}},
  \bibinfo {author} {\bibfnamefont {P.~F.}\ \bibnamefont {Hopkins}}, \ and\
  \bibinfo {author} {\bibfnamefont {A.~C.}\ \bibnamefont {Gossard}},\ }\href
  {\doibase 10.1103/PhysRevLett.69.506} {\bibfield  {journal} {\bibinfo
  {journal} {Phys. Rev. Lett.}\ }\textbf {\bibinfo {volume} {69}},\ \bibinfo
  {pages} {506} (\bibinfo {year} {1992})}\BibitemShut {NoStop}%
\bibitem [{\citenamefont {Haq}\ \emph {et~al.}(1982)\citenamefont {Haq},
  \citenamefont {Pandey},\ and\ \citenamefont {Bohigas}}]{bohigas}%
  \BibitemOpen
  \bibfield  {author} {\bibinfo {author} {\bibfnamefont {R.~U.}\ \bibnamefont
  {Haq}}, \bibinfo {author} {\bibfnamefont {A.}~\bibnamefont {Pandey}}, \ and\
  \bibinfo {author} {\bibfnamefont {O.}~\bibnamefont {Bohigas}},\ }\href
  {\doibase 10.1103/PhysRevLett.48.1086} {\bibfield  {journal} {\bibinfo
  {journal} {Phys. Rev. Lett.}\ }\textbf {\bibinfo {volume} {48}},\ \bibinfo
  {pages} {1086} (\bibinfo {year} {1982})}\BibitemShut {NoStop}%
\bibitem [{\citenamefont {Poilblanc}\ \emph {et~al.}(1993)\citenamefont
  {Poilblanc}, \citenamefont {Ziman}, \citenamefont {Bellissard}, \citenamefont
  {Mila},\ and\ \citenamefont {Montambaux}}]{montambaux}%
  \BibitemOpen
  \bibfield  {author} {\bibinfo {author} {\bibfnamefont {D.}~\bibnamefont
  {Poilblanc}}, \bibinfo {author} {\bibfnamefont {T.}~\bibnamefont {Ziman}},
  \bibinfo {author} {\bibfnamefont {J.}~\bibnamefont {Bellissard}}, \bibinfo
  {author} {\bibfnamefont {F.}~\bibnamefont {Mila}}, \ and\ \bibinfo {author}
  {\bibfnamefont {G.}~\bibnamefont {Montambaux}},\ }\href@noop {} {\bibfield
  {journal} {\bibinfo  {journal} {EPL (Europhysics Letters)}\ }\textbf
  {\bibinfo {volume} {22}},\ \bibinfo {pages} {537} (\bibinfo {year}
  {1993})}\BibitemShut {NoStop}%
\bibitem [{\citenamefont {Mehta}(2004)}]{mehta}%
  \BibitemOpen
  \bibfield  {author} {\bibinfo {author} {\bibfnamefont {M.}~\bibnamefont
  {Mehta}},\ }\href@noop {} {\emph {\bibinfo {title} {Random Matrices}}},\ Pure
  and Applied Mathematics\ (\bibinfo  {publisher} {Elsevier Science},\ \bibinfo
  {year} {2004})\BibitemShut {NoStop}%
\bibitem [{\citenamefont {Haake}(2001)}]{hakke}%
  \BibitemOpen
  \bibfield  {author} {\bibinfo {author} {\bibfnamefont {F.}~\bibnamefont
  {Haake}},\ }\href@noop {} {\emph {\bibinfo {title} {Quantum Signatures of
  Chaos}}},\ Springer Series in Synergetics\ (\bibinfo  {publisher}
  {Springer},\ \bibinfo {year} {2001})\BibitemShut {NoStop}%
\bibitem [{\citenamefont {Rigol}\ \emph {et~al.}(2007)\citenamefont {Rigol},
  \citenamefont {Dunjko}, \citenamefont {Yurovsky},\ and\ \citenamefont
  {Olshanii}}]{gge}%
  \BibitemOpen
  \bibfield  {author} {\bibinfo {author} {\bibfnamefont {M.}~\bibnamefont
  {Rigol}}, \bibinfo {author} {\bibfnamefont {V.}~\bibnamefont {Dunjko}},
  \bibinfo {author} {\bibfnamefont {V.}~\bibnamefont {Yurovsky}}, \ and\
  \bibinfo {author} {\bibfnamefont {M.}~\bibnamefont {Olshanii}},\ }\href
  {\doibase 10.1103/PhysRevLett.98.050405} {\bibfield  {journal} {\bibinfo
  {journal} {Phys. Rev. Lett.}\ }\textbf {\bibinfo {volume} {98}},\ \bibinfo
  {pages} {050405} (\bibinfo {year} {2007})}\BibitemShut {NoStop}%
\bibitem [{\citenamefont {Kollar}\ \emph {et~al.}(2011)\citenamefont {Kollar},
  \citenamefont {Wolf},\ and\ \citenamefont {Eckstein}}]{Eckstein2012}%
  \BibitemOpen
  \bibfield  {author} {\bibinfo {author} {\bibfnamefont {M.}~\bibnamefont
  {Kollar}}, \bibinfo {author} {\bibfnamefont {F.~A.}\ \bibnamefont {Wolf}}, \
  and\ \bibinfo {author} {\bibfnamefont {M.}~\bibnamefont {Eckstein}},\ }\href
  {\doibase 10.1103/PhysRevB.84.054304} {\bibfield  {journal} {\bibinfo
  {journal} {Phys. Rev. B}\ }\textbf {\bibinfo {volume} {84}},\ \bibinfo
  {pages} {054304} (\bibinfo {year} {2011})}\BibitemShut {NoStop}%
\bibitem [{\citenamefont {Cassidy}\ \emph {et~al.}(2011)\citenamefont
  {Cassidy}, \citenamefont {Clark},\ and\ \citenamefont {Rigol}}]{Cassidy2011}%
  \BibitemOpen
  \bibfield  {author} {\bibinfo {author} {\bibfnamefont {A.~C.}\ \bibnamefont
  {Cassidy}}, \bibinfo {author} {\bibfnamefont {C.~W.}\ \bibnamefont {Clark}},
  \ and\ \bibinfo {author} {\bibfnamefont {M.}~\bibnamefont {Rigol}},\ }\href
  {\doibase 10.1103/PhysRevLett.106.140405} {\bibfield  {journal} {\bibinfo
  {journal} {Phys. Rev. Lett.}\ }\textbf {\bibinfo {volume} {106}},\ \bibinfo
  {pages} {140405} (\bibinfo {year} {2011})}\BibitemShut {NoStop}%
\bibitem [{\citenamefont {Steinigeweg}\ \emph {et~al.}(2012)\citenamefont
  {Steinigeweg}, \citenamefont {Herbrych}, \citenamefont
  {Prelov\ifmmode~\check{s}\else \v{s}\fi{}ek},\ and\ \citenamefont
  {Mierzejewski}}]{my3}%
  \BibitemOpen
  \bibfield  {author} {\bibinfo {author} {\bibfnamefont {R.}~\bibnamefont
  {Steinigeweg}}, \bibinfo {author} {\bibfnamefont {J.}~\bibnamefont
  {Herbrych}}, \bibinfo {author} {\bibfnamefont {P.}~\bibnamefont
  {Prelov\ifmmode~\check{s}\else \v{s}\fi{}ek}}, \ and\ \bibinfo {author}
  {\bibfnamefont {M.}~\bibnamefont {Mierzejewski}},\ }\href {\doibase
  10.1103/PhysRevB.85.214409} {\bibfield  {journal} {\bibinfo  {journal} {Phys.
  Rev. B}\ }\textbf {\bibinfo {volume} {85}},\ \bibinfo {pages} {214409}
  (\bibinfo {year} {2012})}\BibitemShut {NoStop}%
\bibitem [{\citenamefont {Zotos}\ and\ \citenamefont
  {Prelov\ifmmode~\check{s}\else \v{s}\fi{}ek}(1996)}]{u2}%
  \BibitemOpen
  \bibfield  {author} {\bibinfo {author} {\bibfnamefont {X.}~\bibnamefont
  {Zotos}}\ and\ \bibinfo {author} {\bibfnamefont {P.}~\bibnamefont
  {Prelov\ifmmode~\check{s}\else \v{s}\fi{}ek}},\ }\href {\doibase
  10.1103/PhysRevB.53.983} {\bibfield  {journal} {\bibinfo  {journal} {Phys.
  Rev. B}\ }\textbf {\bibinfo {volume} {53}},\ \bibinfo {pages} {983} (\bibinfo
  {year} {1996})}\BibitemShut {NoStop}%
\bibitem [{\citenamefont {Mierzejewski}\ and\ \citenamefont
  {Prelov\ifmmode~\check{s}\else \v{s}\fi{}ek}(2010)}]{my1}%
  \BibitemOpen
  \bibfield  {author} {\bibinfo {author} {\bibfnamefont {M.}~\bibnamefont
  {Mierzejewski}}\ and\ \bibinfo {author} {\bibfnamefont {P.}~\bibnamefont
  {Prelov\ifmmode~\check{s}\else \v{s}\fi{}ek}},\ }\href {\doibase
  10.1103/PhysRevLett.105.186405} {\bibfield  {journal} {\bibinfo  {journal}
  {Phys. Rev. Lett.}\ }\textbf {\bibinfo {volume} {105}},\ \bibinfo {pages}
  {186405} (\bibinfo {year} {2010})}\BibitemShut {NoStop}%
\bibitem [{\citenamefont {Mierzejewski}\ \emph {et~al.}(2011)\citenamefont
  {Mierzejewski}, \citenamefont {Bon\ifmmode~\check{c}\else \v{c}\fi{}a},\ and\
  \citenamefont {Prelov\ifmmode~\check{s}\else \v{s}\fi{}ek}}]{my2}%
  \BibitemOpen
  \bibfield  {author} {\bibinfo {author} {\bibfnamefont {M.}~\bibnamefont
  {Mierzejewski}}, \bibinfo {author} {\bibfnamefont {J.}~\bibnamefont
  {Bon\ifmmode~\check{c}\else \v{c}\fi{}a}}, \ and\ \bibinfo {author}
  {\bibfnamefont {P.}~\bibnamefont {Prelov\ifmmode~\check{s}\else
  \v{s}\fi{}ek}},\ }\href {\doibase 10.1103/PhysRevLett.107.126601} {\bibfield
  {journal} {\bibinfo  {journal} {Phys. Rev. Lett.}\ }\textbf {\bibinfo
  {volume} {107}},\ \bibinfo {pages} {126601} (\bibinfo {year}
  {2011})}\BibitemShut {NoStop}%
\bibitem [{\citenamefont {Prosen}(2011)}]{Tomaz2011}%
  \BibitemOpen
  \bibfield  {author} {\bibinfo {author} {\bibfnamefont {T.}~\bibnamefont
  {Prosen}},\ }\href {\doibase 10.1103/PhysRevLett.107.137201} {\bibfield
  {journal} {\bibinfo  {journal} {Phys. Rev. Lett.}\ }\textbf {\bibinfo
  {volume} {107}},\ \bibinfo {pages} {137201} (\bibinfo {year}
  {2011})}\BibitemShut {NoStop}%
\bibitem [{\citenamefont {\ifmmode \check{Z}\else
  \v{Z}\fi{}nidari\ifmmode~\check{c}\else \v{c}\fi{}}(2011)}]{Marko2011}%
  \BibitemOpen
  \bibfield  {author} {\bibinfo {author} {\bibfnamefont {M.}~\bibnamefont
  {\ifmmode \check{Z}\else \v{Z}\fi{}nidari\ifmmode~\check{c}\else
  \v{c}\fi{}}},\ }\href {\doibase 10.1103/PhysRevLett.106.220601} {\bibfield
  {journal} {\bibinfo  {journal} {Phys. Rev. Lett.}\ }\textbf {\bibinfo
  {volume} {106}},\ \bibinfo {pages} {220601} (\bibinfo {year}
  {2011})}\BibitemShut {NoStop}%
\bibitem [{\citenamefont {Sirker}\ \emph {et~al.}(2009)\citenamefont {Sirker},
  \citenamefont {Pereira},\ and\ \citenamefont {Affleck}}]{Sirker2009}%
  \BibitemOpen
  \bibfield  {author} {\bibinfo {author} {\bibfnamefont {J.}~\bibnamefont
  {Sirker}}, \bibinfo {author} {\bibfnamefont {R.~G.}\ \bibnamefont {Pereira}},
  \ and\ \bibinfo {author} {\bibfnamefont {I.}~\bibnamefont {Affleck}},\ }\href
  {\doibase 10.1103/PhysRevLett.103.216602} {\bibfield  {journal} {\bibinfo
  {journal} {Phys. Rev. Lett.}\ }\textbf {\bibinfo {volume} {103}},\ \bibinfo
  {pages} {216602} (\bibinfo {year} {2009})}\BibitemShut {NoStop}%
\bibitem [{\citenamefont {Steinigeweg}\ and\ \citenamefont
  {Brenig}(2011)}]{Robin2011}%
  \BibitemOpen
  \bibfield  {author} {\bibinfo {author} {\bibfnamefont {R.}~\bibnamefont
  {Steinigeweg}}\ and\ \bibinfo {author} {\bibfnamefont {W.}~\bibnamefont
  {Brenig}},\ }\href {\doibase 10.1103/PhysRevLett.107.250602} {\bibfield
  {journal} {\bibinfo  {journal} {Phys. Rev. Lett.}\ }\textbf {\bibinfo
  {volume} {107}},\ \bibinfo {pages} {250602} (\bibinfo {year}
  {2011})}\BibitemShut {NoStop}%
\bibitem [{\citenamefont {Long}\ \emph {et~al.}(2003)\citenamefont {Long},
  \citenamefont {Prelov\ifmmode~\check{s}\else \v{s}\fi{}ek}, \citenamefont
  {El~Shawish}, \citenamefont {Karadamoglou},\ and\ \citenamefont
  {Zotos}}]{mclm}%
  \BibitemOpen
  \bibfield  {author} {\bibinfo {author} {\bibfnamefont {M.~W.}\ \bibnamefont
  {Long}}, \bibinfo {author} {\bibfnamefont {P.}~\bibnamefont
  {Prelov\ifmmode~\check{s}\else \v{s}\fi{}ek}}, \bibinfo {author}
  {\bibfnamefont {S.}~\bibnamefont {El~Shawish}}, \bibinfo {author}
  {\bibfnamefont {J.}~\bibnamefont {Karadamoglou}}, \ and\ \bibinfo {author}
  {\bibfnamefont {X.}~\bibnamefont {Zotos}},\ }\href {\doibase
  10.1103/PhysRevB.68.235106} {\bibfield  {journal} {\bibinfo  {journal} {Phys.
  Rev. B}\ }\textbf {\bibinfo {volume} {68}},\ \bibinfo {pages} {235106}
  (\bibinfo {year} {2003})}\BibitemShut {NoStop}%
\bibitem [{\citenamefont {Goldstein}\ \emph {et~al.}(2006)\citenamefont
  {Goldstein}, \citenamefont {Lebowitz}, \citenamefont {Tumulka},\ and\
  \citenamefont {Zangh\`\i}}]{Goldstein2006}%
  \BibitemOpen
  \bibfield  {author} {\bibinfo {author} {\bibfnamefont {S.}~\bibnamefont
  {Goldstein}}, \bibinfo {author} {\bibfnamefont {J.~L.}\ \bibnamefont
  {Lebowitz}}, \bibinfo {author} {\bibfnamefont {R.}~\bibnamefont {Tumulka}}, \
  and\ \bibinfo {author} {\bibfnamefont {N.}~\bibnamefont {Zangh\`\i}},\ }\href
  {\doibase 10.1103/PhysRevLett.96.050403} {\bibfield  {journal} {\bibinfo
  {journal} {Phys. Rev. Lett.}\ }\textbf {\bibinfo {volume} {96}},\ \bibinfo
  {pages} {050403} (\bibinfo {year} {2006})}\BibitemShut {NoStop}%
\bibitem [{\citenamefont {Park}\ and\ \citenamefont {Light}(1986)}]{lantime}%
  \BibitemOpen
  \bibfield  {author} {\bibinfo {author} {\bibfnamefont {T.~J.}\ \bibnamefont
  {Park}}\ and\ \bibinfo {author} {\bibfnamefont {J.~C.}\ \bibnamefont
  {Light}},\ }\href {\doibase 10.1063/1.451548} {\bibfield  {journal} {\bibinfo
   {journal} {The Journal of Chemical Physics}\ }\textbf {\bibinfo {volume}
  {85}},\ \bibinfo {pages} {5870} (\bibinfo {year} {1986})}\BibitemShut
  {NoStop}%
\bibitem [{\citenamefont {Bruus}\ and\ \citenamefont
  {Angl`es~d'Auriac}(1997)}]{Bruus1997}%
  \BibitemOpen
  \bibfield  {author} {\bibinfo {author} {\bibfnamefont {H.}~\bibnamefont
  {Bruus}}\ and\ \bibinfo {author} {\bibfnamefont {J.-C.}\ \bibnamefont
  {Angl`es~d'Auriac}},\ }\href {\doibase 10.1103/PhysRevB.55.9142} {\bibfield
  {journal} {\bibinfo  {journal} {Phys. Rev. B}\ }\textbf {\bibinfo {volume}
  {55}},\ \bibinfo {pages} {9142} (\bibinfo {year} {1997})}\BibitemShut
  {NoStop}%
\bibitem [{\citenamefont {Brody}\ \emph {et~al.}(1981)\citenamefont {Brody},
  \citenamefont {Flores}, \citenamefont {French}, \citenamefont {Mello},
  \citenamefont {Pandey},\ and\ \citenamefont {Wong}}]{Brody1981}%
  \BibitemOpen
  \bibfield  {author} {\bibinfo {author} {\bibfnamefont {T.~A.}\ \bibnamefont
  {Brody}}, \bibinfo {author} {\bibfnamefont {J.}~\bibnamefont {Flores}},
  \bibinfo {author} {\bibfnamefont {J.~B.}\ \bibnamefont {French}}, \bibinfo
  {author} {\bibfnamefont {P.~A.}\ \bibnamefont {Mello}}, \bibinfo {author}
  {\bibfnamefont {A.}~\bibnamefont {Pandey}}, \ and\ \bibinfo {author}
  {\bibfnamefont {S.~S.~M.}\ \bibnamefont {Wong}},\ }\href {\doibase
  10.1103/RevModPhys.53.385} {\bibfield  {journal} {\bibinfo  {journal} {Rev.
  Mod. Phys.}\ }\textbf {\bibinfo {volume} {53}},\ \bibinfo {pages} {385}
  (\bibinfo {year} {1981})}\BibitemShut {NoStop}%
\bibitem [{\citenamefont {Guhr}\ \emph {et~al.}(1998)\citenamefont {Guhr},
  \citenamefont {Muller-Groeling},\ and\ \citenamefont
  {Weidenmuller}}]{Guhr1998}%
  \BibitemOpen
  \bibfield  {author} {\bibinfo {author} {\bibfnamefont {T.}~\bibnamefont
  {Guhr}}, \bibinfo {author} {\bibfnamefont {A.}~\bibnamefont
  {Muller-Groeling}}, \ and\ \bibinfo {author} {\bibfnamefont {H.~A.}\
  \bibnamefont {Weidenmuller}},\ }\href {\doibase
  10.1016/S0370-1573(97)00088-4} {\bibfield  {journal} {\bibinfo  {journal}
  {Physics Reports}\ }\textbf {\bibinfo {volume} {299}},\ \bibinfo {pages} {189
  } (\bibinfo {year} {1998})}\BibitemShut {NoStop}%
\bibitem [{\citenamefont {Rigol}\ and\ \citenamefont
  {Srednicki}(2012)}]{Rigol2012}%
  \BibitemOpen
  \bibfield  {author} {\bibinfo {author} {\bibfnamefont {M.}~\bibnamefont
  {Rigol}}\ and\ \bibinfo {author} {\bibfnamefont {M.}~\bibnamefont
  {Srednicki}},\ }\href {\doibase 10.1103/PhysRevLett.108.110601} {\bibfield
  {journal} {\bibinfo  {journal} {Phys. Rev. Lett.}\ }\textbf {\bibinfo
  {volume} {108}},\ \bibinfo {pages} {110601} (\bibinfo {year}
  {2012})}\BibitemShut {NoStop}%
\bibitem [{\citenamefont {Santos}\ \emph {et~al.}(2012)\citenamefont {Santos},
  \citenamefont {Polkovnikov},\ and\ \citenamefont {Rigol}}]{rigolast}%
  \BibitemOpen
  \bibfield  {author} {\bibinfo {author} {\bibfnamefont {L.~F.}\ \bibnamefont
  {Santos}}, \bibinfo {author} {\bibfnamefont {A.}~\bibnamefont {Polkovnikov}},
  \ and\ \bibinfo {author} {\bibfnamefont {M.}~\bibnamefont {Rigol}},\ }\href
  {\doibase 10.1103/PhysRevE.86.010102} {\bibfield  {journal} {\bibinfo
  {journal} {Phys. Rev. E}\ }\textbf {\bibinfo {volume} {86}},\ \bibinfo
  {pages} {010102} (\bibinfo {year} {2012})}\BibitemShut {NoStop}%
\bibitem [{\citenamefont {Eisert}\ \emph {et~al.}(2010)\citenamefont {Eisert},
  \citenamefont {Cramer},\ and\ \citenamefont {Plenio}}]{Eisert2010}%
  \BibitemOpen
  \bibfield  {author} {\bibinfo {author} {\bibfnamefont {J.}~\bibnamefont
  {Eisert}}, \bibinfo {author} {\bibfnamefont {M.}~\bibnamefont {Cramer}}, \
  and\ \bibinfo {author} {\bibfnamefont {M.~B.}\ \bibnamefont {Plenio}},\
  }\href {\doibase 10.1103/RevModPhys.82.277} {\bibfield  {journal} {\bibinfo
  {journal} {Rev. Mod. Phys.}\ }\textbf {\bibinfo {volume} {82}},\ \bibinfo
  {pages} {277} (\bibinfo {year} {2010})}\BibitemShut {NoStop}%
\bibitem [{\citenamefont {Manmana}\ \emph {et~al.}(2007)\citenamefont
  {Manmana}, \citenamefont {Wessel}, \citenamefont {Noack},\ and\ \citenamefont
  {Muramatsu}}]{Manmana2007}%
  \BibitemOpen
  \bibfield  {author} {\bibinfo {author} {\bibfnamefont {S.~R.}\ \bibnamefont
  {Manmana}}, \bibinfo {author} {\bibfnamefont {S.}~\bibnamefont {Wessel}},
  \bibinfo {author} {\bibfnamefont {R.~M.}\ \bibnamefont {Noack}}, \ and\
  \bibinfo {author} {\bibfnamefont {A.}~\bibnamefont {Muramatsu}},\ }\href
  {\doibase 10.1103/PhysRevLett.98.210405} {\bibfield  {journal} {\bibinfo
  {journal} {Phys. Rev. Lett.}\ }\textbf {\bibinfo {volume} {98}},\ \bibinfo
  {pages} {210405} (\bibinfo {year} {2007})}\BibitemShut {NoStop}%
\bibitem [{\citenamefont {Gogolin}\ \emph {et~al.}(2011)\citenamefont
  {Gogolin}, \citenamefont {M\"uller},\ and\ \citenamefont
  {Eisert}}]{Gogolin2011}%
  \BibitemOpen
  \bibfield  {author} {\bibinfo {author} {\bibfnamefont {C.}~\bibnamefont
  {Gogolin}}, \bibinfo {author} {\bibfnamefont {M.~P.}\ \bibnamefont
  {M\"uller}}, \ and\ \bibinfo {author} {\bibfnamefont {J.}~\bibnamefont
  {Eisert}},\ }\href {\doibase 10.1103/PhysRevLett.106.040401} {\bibfield
  {journal} {\bibinfo  {journal} {Phys. Rev. Lett.}\ }\textbf {\bibinfo
  {volume} {106}},\ \bibinfo {pages} {040401} (\bibinfo {year}
  {2011})}\BibitemShut {NoStop}%
\end{thebibliography}%

\end{document}